\documentclass{PoS}

\newcommand{\bee}[0]{\begin{eqnarray}}
\newcommand{\eee}[0]{\end{eqnarray}}

\title{Leptogenesis in natural low-scale seesaw mechanisms}

\ShortTitle{Leptogenesis in natural low-scale seesaw mechanisms}

\author{\speaker{Michele Lucente}
        \thanks{In collaboration with Asm\^aa Abada, Giorgio Arcadi and Valerie Domcke.}\\
        \\
       Laboratoire de Physique Th\'eorique, CNRS -- UMR 8627, Universit\'e de Paris-Sud\\
       F-91405 Orsay Cedex, France\\
       \\
       Scuola Internazionale Superiore di Studi Avanzati\\
       Via Bonomea 265, 34136 Trieste, Italy\\
       \\
       E-mail: \email{mlucente@sissa.it}}

\abstract{We explore the hypothesis of having an approximate lepton number conservation as a way to achieve a successful leptogenesis in low-scale seesaw mechanisms. The smallness of the active neutrino masses, as well as a strong degeneracy in the mass spectrum of the heavy sterile states, are both consequence of the assumed approximate symmetry. We propose a minimal extension of the Standard Model in order to implement the idea, and perform an analytical and numerical study to determine the viable solutions in the model and the testability of this leptogenesis scenario in future experiments.
\\
\\
\\
\\

\hfill LPT-Orsay-15-72\\[-1\baselineskip]

\hfill SISSA  48/2015/FISI
}

\FullConference{The European Physical Society Conference on High Energy Physics\\
		 22-29 July 2015\\
		 Vienna, Austria}

\begin{document}

\section{Introduction}
The massive nature of neutrinos and the baryon asymmetry of the Universe (BAU) are two observations that call for an extension of the Standard Model of particle physics (SM). A popular and common solution to both these issues is the introduction of fermionic fields that are singlet under the SM gauge group: these fields can account for neutrino masses via a seesaw mechanism, while the heavy eigenstates in the mass spectrum (sterile neutrinos) can  generate a lepton asymmetry in the early Universe while they decay out of equilibrium (thermal leptogenesis)~\cite{Davidson:2008bu}. The lepton asymmetry is eventually converted into a baryon asymmetry by sphalerons. In order to account for the observed BAU via thermal leptogenesis the extra SM states must in general have masses larger than $10^8$ GeV, making this mechanism impossible to probe in current experimental facilities. The value of the new physics (NP) scale can be nonetheless lower ($\sim$ TeV) in the presence of a resonant amplification of the lepton asymmetry, if the new states feature a degeneracy in their mass spectrum~\cite{Pilaftsis:2003gt}. At even lower scales ($\sim$ GeV) thermal leptogenesis is no more effective, but a sizeable lepton asymmetry can be nonetheless generated during the production of the heavy states in the early Universe~\cite{Akhmedov:1998qx,Asaka:2005pn,Shaposhnikov:2008pf,Asaka:2010kk,Asaka:2011wq,Canetti:2012zc,Canetti:2012kh}. This leptogenesis mechanism is very interesting since the new states lie at the GeV scale and can thus be probed in future experiments.

\section{Leptogenesis, neutrino masses and lepton number violation}
The minimal setup for a viable leptogenesis at the GeV scale requires the existence of a pair of sterile neutrinos which are strongly degenerate in mass. These states oscillate among themselves, generating an asymmetry in their sterile flavour numbers. An asymmetry in the active flavours is then generated, given the flavour violating Yukawa interactions that connect the sterile neutrinos to the Higgs and the active lepton doublets. Notice however that due to the small value of the ratio between the Majorana masses of the sterile neutrinos, $M \sim$ GeV, and the temperature at which the leptogenesis process takes place $T\gtrsim T_{W} \approx 140$ GeV,\footnote{$T_W$ is the temperature of the electroweak phase transition.} the rates of the lepton number violating interactions can be safely neglected. The total lepton asymmetry, defined as the sum of the asymmetries in the individual (active and sterile) flavours, is thus vanishing. However, since sphalerons only couple to the active leptons, they eventually convert the asymmetry in the active flavours (and only that asymmetry) into a net baryon asymmetry.

\begin{figure}[htb]
 \begin{tabular}{cc}
\includegraphics[width=0.45\textwidth]{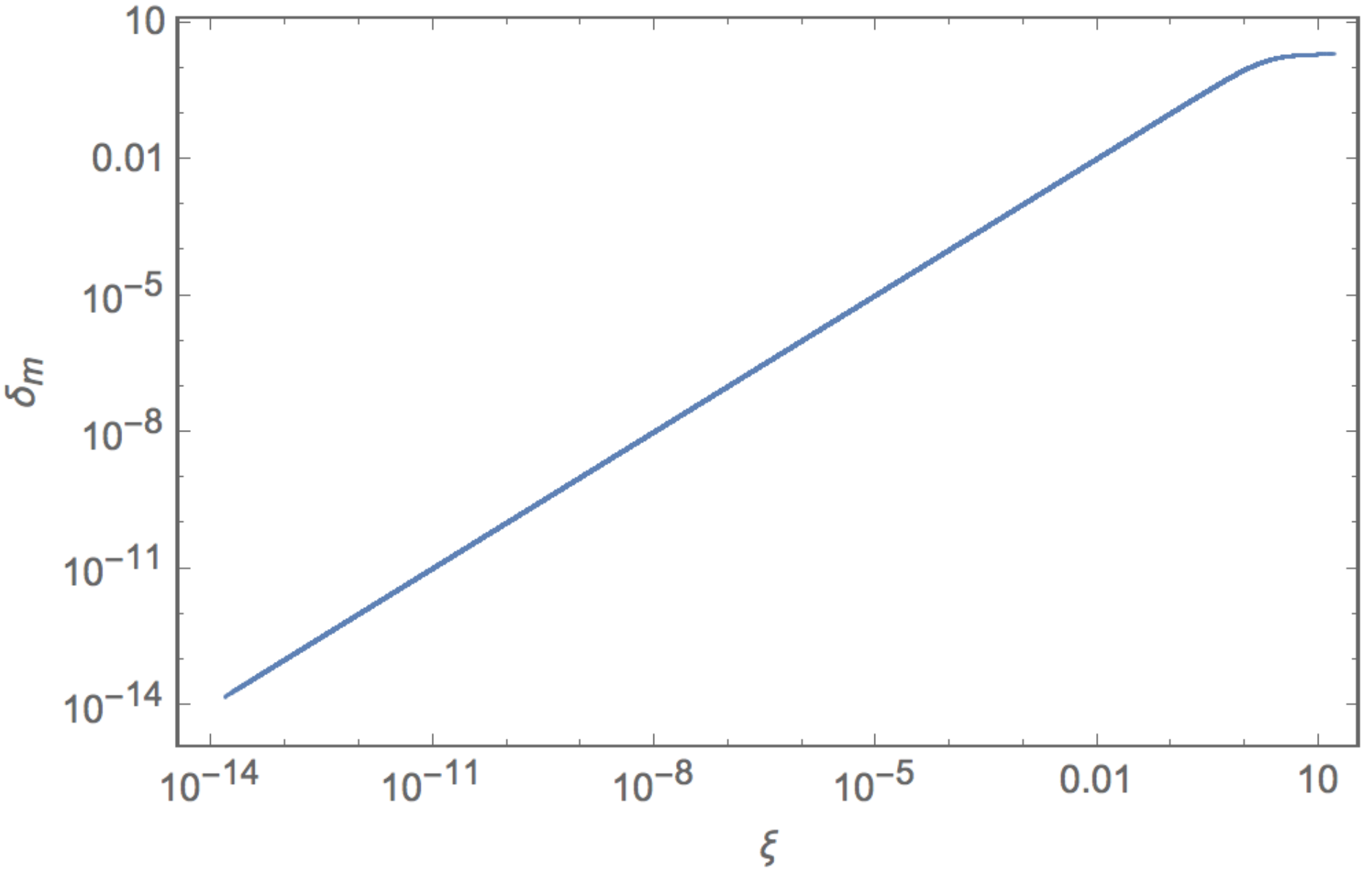}
\hspace*{2mm}&\hspace*{2mm}
\includegraphics[width=0.45\textwidth]{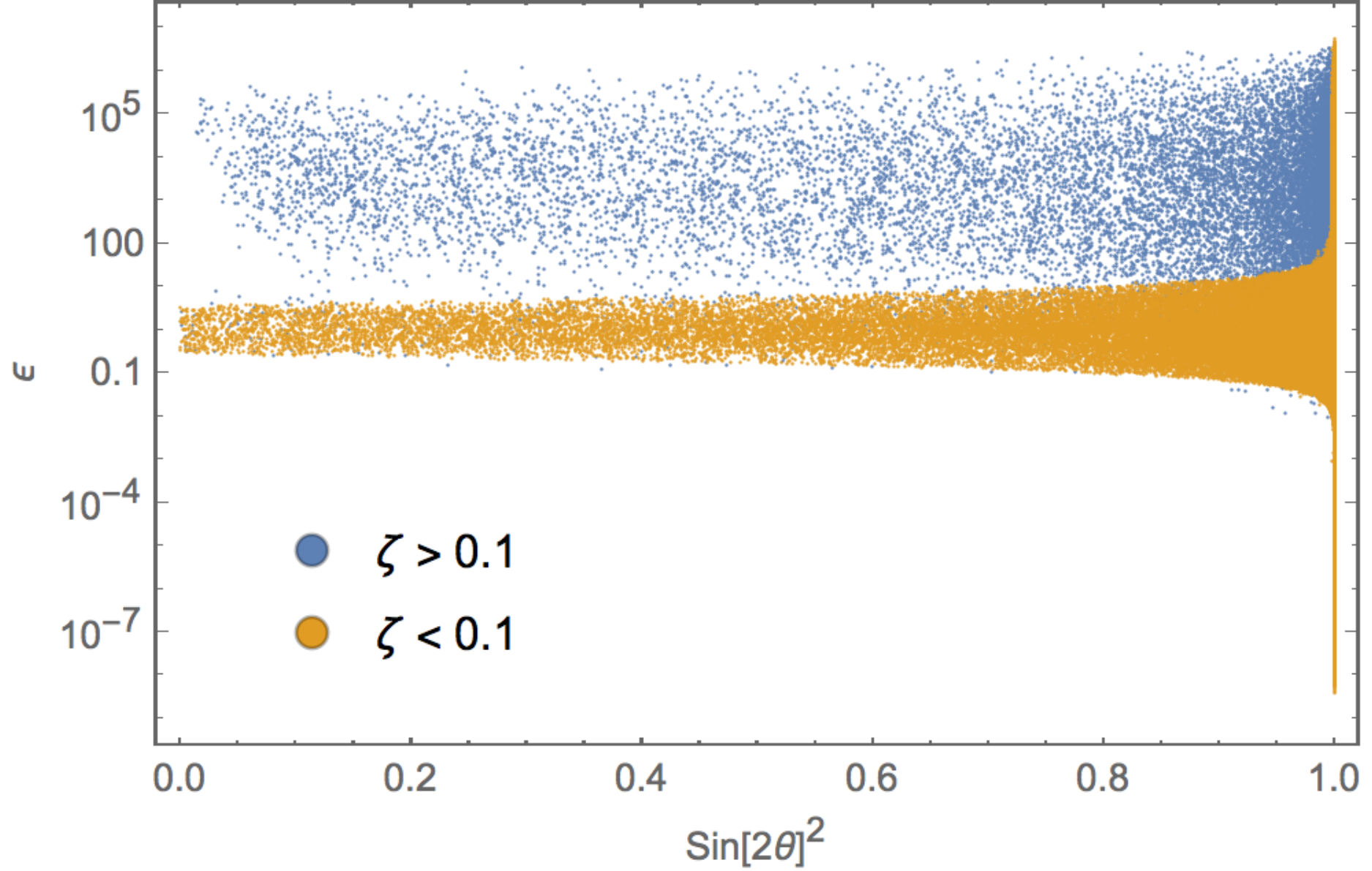}
\end{tabular}
\caption{Correlation between the lepton number violating parameters in the model and the quantities relevant for the oscillation dynamics in the heavy neutrino sector. \emph{Left panel}: relative mass splitting $\delta_m \equiv 2 (M_5-M_4)/(M_5+M_4)$ as a function of the ``ISS'' perturbation $\xi$. \emph{Right panel}: mixing angle as a function of the ``LSS'' perturbation $\epsilon$; blue (orange) points refer to solutions with $\xi>0.1$ ($\xi<0.1$).}
\label{fig:param_osc}
\end{figure}

Notice that the basic ingredients for a successful leptogenesis at the GeV scale (low-scale seesaw and a strong degeneracy in the masses of the sterile sector) are naturally present in neutrino mass generation mechanisms characterised by an approximate  conservation of the total lepton number $L$. In this framework the scale of the (lepton number violating) active neutrino Majorana masses is suppressed by the approximate symmetry, while the heavy states couple to form pseudo-Dirac pairs which are strongly degenerate in mass. The simplest implementation of this idea consists in extending the SM by adding two sterile fermions $N_R^{i}$ with opposite lepton number, $L=\pm 1$. Considering for simplicity a toy model with only one active flavour $\nu_L$, the lepton conserving part of the mass matrix is, in the basis $(\nu_L, {N_R^{1}}^c, {N_R^{2}}^c)$,
\bee\label{eq:M0_toy}
M_0 &=& \left(\begin{array}{ccc} 0 & \frac{1}{\sqrt{2}}Y v & 0\\
\frac{1}{\sqrt{2}} Y v & 0 & \Lambda \\
0 & \Lambda & 0
\end{array}\right),
\eee
where $v$ denotes the Higgs vacuum expectation value ($v=246$ GeV at zero temperature), $Y$ is the Yukawa coupling between the sterile fermion with $L=1$ and the Higgs and lepton doublets, $\Lambda$ is a mass parameter coupling the sterile fermions. The mass spectrum resulting from this mass matrix is
\begin{equation}
m_\nu =0,\hspace{1cm}
M_{1,2} = \sqrt{|\Lambda|^2 +\frac{1}{2}|Y v|^2}.
\label{eq_M}
\end{equation}
In order to account for both nonzero neutrino masses and a viable leptogenesis it is necessary to perturb this mass matrix by inserting small lepton number violating parameters, which will lift the degeneracy in the heavy states and generate neutrino masses. A term in the (1,1) entry of $M_0$ violates gauge invariance and can only be generated in non-minimal models, for instance by adding an Higgs isospin triplet. We are not interested in such a case, and there are thus 3 possible ways to perturb $M_0$
\begin{equation}
\begin{array}{lcccr}
\Delta M_{ISS}=\left(\begin{array}{ccc}
 0 & 0 & 0\\
0 & 0 & 0\\
0 & 0 & \xi \Lambda
\end{array}\right),& \hspace{0.01\textwidth} 
&
\Delta M_{LSS}=\left( \begin{array}{ccc} 0 & 0 & \frac{\epsilon}{\sqrt{2}} Y' v\\
0 & 0 & 0\\
\frac{\epsilon}{\sqrt{2}} \, Y' v & 0 & 0 
\end{array}\right),& \hspace{0.01\textwidth} 
&
\Delta M_{ESS}=\left(\begin{array}{ccc} 0 & 0 & 0\\
0 & \xi' \Lambda & 0\\
0 & 0 & 0 
\end{array}\right),
\end{array}
\label{eq_DMs}
\end{equation}
where $\xi, \epsilon$ and $\xi'$ are small dimensionless parameters. The first pattern reproduces the Inverse Seesaw (ISS) mechanisms~\cite{Wyler:1982dd,Mohapatra:1986bd}, the second one the Linear Seesaw (LSS)~\cite{Barr:2003nn,Malinsky:2005bi} while the third one is the Extended Seesaw (ESS)~\cite{Kang:2006sn} which generates neutrino masses at loop level. By imposing the requirement of reproducing the observed values for the active neutrino masses, $m_\nu \gtrsim \sqrt{\Delta m_\textrm{atm}^2}\simeq 5 \times 10^{-2}$ eV, and the condition that the Yukawa couplings for the heavy states are small enough such that they remain out of equilibrium during the leptogenesis process, $Y < \sqrt{2} \times 10^{-7}$~\cite{Akhmedov:1998qx}, it is possible to show that the ISS pattern provides a mass spitting in the heavy sector which is too large in order to account for a successful leptogenesis~\cite{Abada:2015rta}. On the other hand the mass splitting generated by the LSS pattern depends on the Higgs VEV $v$, and it thus vanishes before the electroweak phase transition, when the leptogenesis process is effective. The ESS suffers from the same problems of the ISS but with a still larger mass splitting. We thus conclude that the minimal framework in order to naturally account for both neutrino masses and a low-scale leptogenesis, based on an approximate lepton number conservation,  is the SM extended by two sterile fermions with an opposite lepton number assignment, considering both an ISS and a LSS-like perturbations. The mass matrix for this model in the realistic 3-flavour case is given by (cf. also~\cite{Gavela:2009cd})
\begin{equation}\label{eq_Mpertexp}
 \mathcal{M} =  \left( \begin{array}{ccc}
 \mathbf{0} &\frac{1}{\sqrt{2}} \textbf{Y} v &\frac{\epsilon}{\sqrt{2}} \textbf{Y}' v\\
 \frac{1}{\sqrt{2}} \textbf{Y}^T v & 0 & \Lambda \\
 \frac{\epsilon}{\sqrt{2}} {\textbf{Y}'}^T v & \Lambda & \xi \Lambda 
 \end{array}
\right),
\end{equation}
where the notation is a self-evident generalisation of the one in eqs.~(\ref{eq:M0_toy}, \ref{eq_DMs}), with the 3-dimensional vectors $\textbf{Y}, \textbf{Y}'$ containing the Yukawa couplings which are all assumed to be of the same order of magnitude.
Notice that the ordering of the second and third column/row  of Eqs.~(\ref{eq:M0_toy}, \ref{eq_Mpertexp}) arises from the lepton number assignment $L=+1$ and $-1$, respectively. Choosing $\epsilon > 1$ and $|\textbf{Y}| \simeq |\textbf{Y}'|$ correspondingly smaller, implies switching this assignment. Thus very large values of $\epsilon \gg 1$ also correspond to an approximate conservation of the lepton number, and there is an approximate symmetry under $\epsilon \rightarrow 1/\epsilon$ which becomes exact when $\xi, \xi' \rightarrow 0$.
It is interesting that the small lepton number violating parameters in eq.~(\ref{eq_Mpertexp}) are directly related to the oscillation dynamics among the two heavy states: the parameter $\xi$ determines the relative mass splitting, while the parameter $\epsilon$ is related to the mixing among the states, cf. Fig.~\ref{fig:param_osc}.

\section{Viable parameter space and testability}
\begin{figure}[htb]
 \begin{tabular}{cc}
\includegraphics[width=0.45\textwidth]{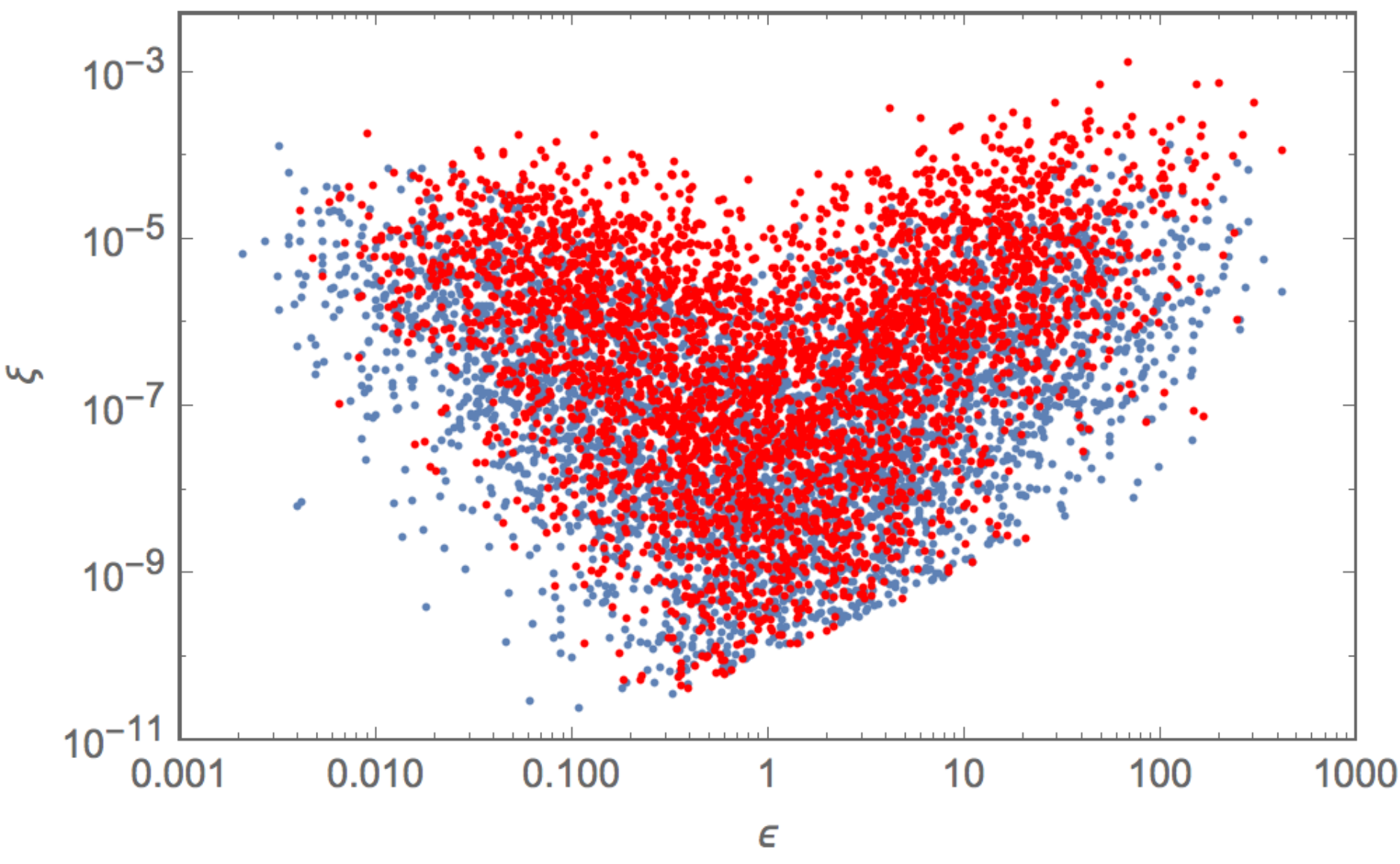}
\hspace*{2mm}&\hspace*{2mm}
\includegraphics[width=0.45\textwidth]{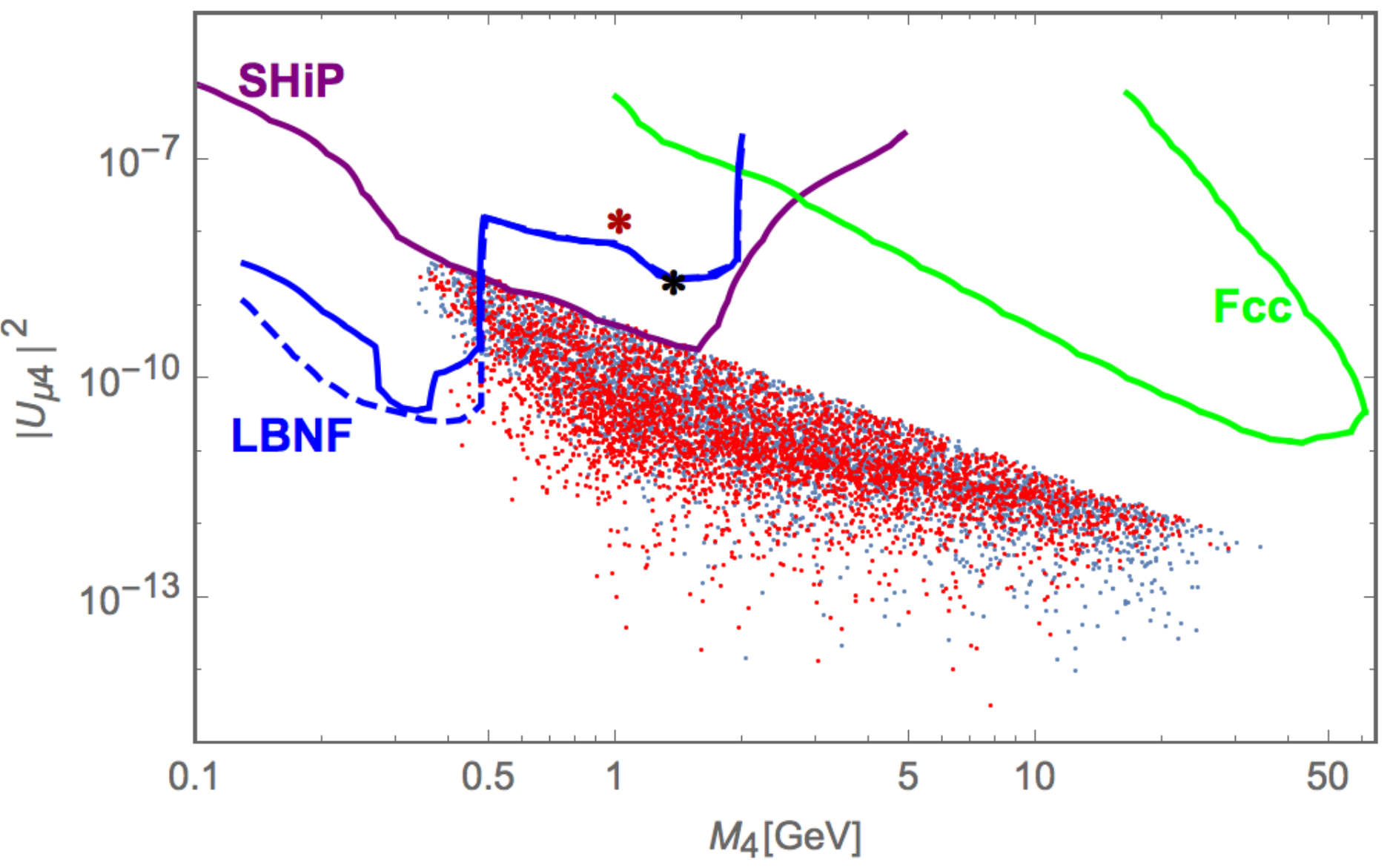}
\end{tabular}
\caption{Viable solutions accounting for both neutrino masses (and mixing) and leptogenesis in the weak washout regime. Blue (red) points refer to a normal (inverse) ordering for the light neutrino mass spectrum. \emph{Left panel}: viable parameter space for the lepton number violating parameters $\xi,\epsilon$. \emph{Right panel}: viable solutions in the plane ($|U_{\mu 4}|^2$, $M_4$), where $U_{\mu 4}$ is the active-sterile mixing in the $\mu$ flavour for the lightest sterile neutrino with mass $M_4$; the expected sensitivity of planned future experiments is reported for comparison. The asterisks refer to two benchmark solutions in the strong washout regime.}
\label{fig:param_pert}
\end{figure}

In the regime of weak washout for the generation of a lepton asymmetry, defined as the one in which the heavy sterile neutrinos are far from thermal equilibrium for the entire duration of the leptogenesis process, the final baryon asymmetry is given by~\cite{Abada:2015rta}
\begin{equation}
\label{eq:baryo_analytical}
Y_{\Delta B}=\frac{n_{\Delta B}}{s}= \frac{945}{2528} \frac{\, 2^{2/3}}{  \,\,  3^{1/3} \, \pi^{5/2}  \,  \Gamma(5/6)} \frac{1}{g_s}\sin^3 \phi \, \frac{M_0}{T_{\rm W}} \frac{M_0^{4/3}}{ \left(\Delta m^2\right)^{2/3}} \, Tr\left[ F^\dagger \delta F\right] \ ,
\end{equation}
where $F_{\alpha j}  = Y_{\alpha I} U_{I j}$ are the Yukawa couplings in the mass basis, $U$ is the leptonic mixing matrix and the indices run over $\alpha = e,\mu,\tau$ (active flavours), $I=1,2$ (sterile flavours) and $j=1,\dots, 5$ (mass eigenstates). 
 $\Delta m^2=M_5^2-M_4^2$ is the mass squared difference of the heavy neutrinos, $g_s$ represents the degrees of freedom in the thermal bath at $T = T_{W}$, $M_0 \approx 7 \times 10^{17}\,\mbox{GeV}$, $\sin\phi \sim 0.012$ and $\delta = \textrm{diag}(\delta_\alpha)$ is defined as:
\begin{equation}\label{eq:deltaCP}
\delta_{\alpha}=\sum_{i>j} \textrm{Im}\left[F_{\alpha i} \left(F^{\dagger} F\right)_{ij} F^{\dagger}_{j\alpha}\right]\ .
\end{equation}
As a rule of thumb, the weak washout regime can be defined by the condition $\left|F_{\alpha j} \right| < \sqrt{2} \times 10^{-7}$ for any combination of indices $\alpha,j$. By means of the expression (\ref{eq:baryo_analytical}) we have performed a scan of the parameter space of the model, in order to determine the viable parameter space that accounts for both the observed neutrino masses (and mixing) and for a viable leptogenesis in the weak washout regime. The results of the scan are reported in Fig.~\ref{fig:param_pert}: the left panel reports the viable parameter space for the lepton number violating parameters, $\epsilon$ and $\xi$, while the right panel reports the active-sterile mixing of the solutions in the $\mu$ flavour, as a function of the mass of the lightest sterile neutrino in the heavy pair, together with the expected sensitivity of some planned experiments~\cite{Adams:2013qkq,Alekhin:2015byh}.
In order to achieve a successful leptogenesis in the weak washout regime a strong degeneracy in the masses of the heavy pair, of the order of $10^{-3}$ or stronger, is needed. As it is evident from the right panel in Fig.~\ref{fig:param_pert}, the requirement of the sterile neutrinos being out of thermal equilibrium puts an upper bound on the allowed active-sterile mixing, with the result that the viable solutions in the weak washout regime are not testable in future facilities, apart from a small region of very light masses that can be probed by LBFN/DUNE. This conclusion, however, does not prevent the model to be tested in future experiments, since the above described analysis has been based on eq.~(\ref{eq:baryo_analytical}) which, we recall, is valid only in the weak washout regime. Viable solutions are present for larger values of the Yukawa couplings as well, and consequently for larger values of the active-sterile mixing. In this regime (strong washout) the heavy neutrinos equilibrate at late times, washing out the produced lepton asymmetry; if the depletion is not complete a residual lepton asymmetry is nonetheless present at the time of the electroweak phase transition, when the sphaleron transitions freeze-out a non-zero residual baryon asymmetry. Since the analytic expression in eq.~(\ref{eq:baryo_analytical}) is not valid in this regime, it is necessary to rely on numerical  methods. The computation is numerically challenging and a complete scan of the parameter space is beyond the scope of the work. Our analysis has been based on the numerical study of a set of benchmark points, in order to demonstrate the existence and testability of these solutions, two of which are reported in the right panel of Fig.~\ref{fig:param_pert} as asterisks. As it is evident, solutions in the strong washout regime are within the testability of future experiments.

\section{Conclusion}
In this work we have proposed and implemented the hypothesis of having an approximate lepton number conservation as a key to achieve leptogenesis in low-scale seesaw mechanisms ($\sim$ GeV). This approximate symmetry accounts at the same time for the smallness of the observed neutrino masses, for the low value of the new physics scale and for the strong degeneracy in the mass spectrum of the heavy sterile neutrinos, which in turn induces an asymmetry in the individual flavour numbers due to CP-violating oscillations. We have proposed a minimal extension of the SM in order to implement this hypothesis, and studied its solutions by means of an analytical expression for the final baryon asymmetry valid in the weak washout regime, which has been validated and complemented by the numerical solution of a set of benchmark points in both the weak and strong washout regimes. We have found that, although the solutions in the weak washout regime are outside the sensitivity of future experiments, solutions in the strong washout regime are testable in planned laboratory facilities.

\section*{Acknowledgments}
We acknowledge support from the European Union FP7  ITN-INVISIBLES (Marie Curie Actions, PITN-GA-2011-289442).


\begin{thebibliography}{99}
\bibitem{Davidson:2008bu}
  S.~Davidson, E.~Nardi and Y.~Nir,
  Phys.\ Rept.\  {\bf 466} (2008) 105
  [arXiv:0802.2962 [hep-ph]].
  
\bibitem{Pilaftsis:2003gt}
  A.~Pilaftsis and T.~E.~J.~Underwood,
  Nucl.\ Phys.\ B {\bf 692} (2004) 303
  [hep-ph/0309342].
  
\bibitem{Akhmedov:1998qx}
  E.~K.~Akhmedov, V.~A.~Rubakov and A.~Y.~Smirnov,
  Phys.\ Rev.\ Lett.\  {\bf 81} (1998) 1359
  [hep-ph/9803255].
  
\bibitem{Asaka:2005pn}
  T.~Asaka and M.~Shaposhnikov,
  Phys.\ Lett.\ B {\bf 620} (2005) 17
  [hep-ph/0505013].
  
\bibitem{Shaposhnikov:2008pf}
  M.~Shaposhnikov,
  JHEP {\bf 0808} (2008) 008
  [arXiv:0804.4542 [hep-ph]].
  
\bibitem{Asaka:2010kk}
  T.~Asaka and H.~Ishida,
  Phys.\ Lett.\ B {\bf 692} (2010) 105
  [arXiv:1004.5491 [hep-ph]].
  
\bibitem{Asaka:2011wq}
  T.~Asaka, S.~Eijima and H.~Ishida,
  JCAP {\bf 1202} (2012) 021
  [arXiv:1112.5565 [hep-ph]].
  
\bibitem{Canetti:2012zc}
  L.~Canetti, M.~Drewes and M.~Shaposhnikov,
  New J.\ Phys.\  {\bf 14} (2012) 095012
  [arXiv:1204.4186 [hep-ph]].
  
\bibitem{Canetti:2012kh}
  L.~Canetti, M.~Drewes, T.~Frossard and M.~Shaposhnikov,
  Phys.\ Rev.\ D {\bf 87} (2013) 9,  093006
  [arXiv:1208.4607 [hep-ph]].
  
\bibitem{Wyler:1982dd}
  D.~Wyler and L.~Wolfenstein,
  Nucl.\ Phys.\ B {\bf 218} (1983) 205.
  
\bibitem{Mohapatra:1986bd}
  R.~N.~Mohapatra and J.~W.~F.~Valle,
  Phys.\ Rev.\ D {\bf 34} (1986) 1642.
  
\bibitem{Barr:2003nn}
  S.~M.~Barr,
  Phys.\ Rev.\ Lett.\  {\bf 92} (2004) 101601
  [hep-ph/0309152].
  
\bibitem{Malinsky:2005bi}
  M.~Malinsky, J.~C.~Romao and J.~W.~F.~Valle,
  Phys.\ Rev.\ Lett.\  {\bf 95} (2005) 161801
  [hep-ph/0506296].
  
\bibitem{Kang:2006sn}
  S.~K.~Kang and C.~S.~Kim,
  Phys.\ Lett.\ B {\bf 646} (2007) 248
  [hep-ph/0607072].
  
\bibitem{Abada:2015rta}
  A.~Abada, G.~Arcadi, V.~Domcke and M.~Lucente,
  arXiv:1507.06215 [hep-ph].
  
\bibitem{Gavela:2009cd}
  M.~B.~Gavela, T.~Hambye, D.~Hernandez and P.~Hernandez,
  JHEP {\bf 0909} (2009) 038
  [arXiv:0906.1461 [hep-ph]].
  
\bibitem{Adams:2013qkq}
  C.~Adams {\it et al.} [LBNE Collaboration],
  arXiv:1307.7335 [hep-ex].
  
\bibitem{Alekhin:2015byh}
  S.~Alekhin {\it et al.},
  arXiv:1504.04855 [hep-ph].
\end{thebibliography}
\end{document}